\documentclass{article}
\usepackage{graphicx}
\title{Automated Alignment of Imperfect EM Images for Neural Reconstruction}
\begin{document}
\author{Louis K. Scheffer, Bill Karsh, and Shiv Vitaladevuni\thanks{Lou Scheffer and Bill Karsh are with Janelia Farms Research Campus,
Howard Hughes Medical Institute.  Shiv Vitaladevuni is with BBN.  Email: SchefferL@janelia.hhmi.org, Karshb@janelia.hhmi.org, SVitalad@bbn.com }}
\date{\today}
\maketitle
\begin{abstract}
The most established method of reconstructing neural circuits from animals involves slicing tissue very thin, then taking mosaics of electron microscope (EM) images.
To trace neurons across different images and through different sections, these images must be accurately aligned, 
both with the others in the same section and to the sections above and below.  Unfortunately, sectioning and imaging are not ideal processes - 
some of the problems that make alignment difficult include lens distortion, 
tissue shrinkage during imaging, tears and folds in the sectioned tissue, and dust and other artifacts.
In addition the data sets are large (hundreds of thousands of images) and each image must be aligned with many neighbors, so the process must be
automated and reliable.  
This paper discusses methods of dealing with these problems, with numeric results describing the accuracy of
the resulting alignments.
\end{abstract}
\section{Introduction}
The most time-tested method of reconstructing networks of neurons involves slicing tissue very thin (sectioning), 
and then taking mosaics of EM images of each section.\cite{chklovskii2010semi}
These images must be aligned before either humans or computers can trace neurons across the images and sections. 
This alignment can be done manually with machine assistance\cite{carlbom1994computer} or 
automatically\cite{tasdizen2010automatic}.
However, given the delicacy of the thin slices, it is extremely difficult to get 1000+ consecutive sections where
every image is perfect.
Therefore, in practice, every such stack includes a number of image imperfections that make alignment non-trivial.
These include contrast and brightness variations, folds, tears, image shrinkage during imaging,
lens distortion, and dust particles, among other problems.  
Yet another challenge for alignment is that even in the best of images,
adjacent sections have features that are merely similar and not identical, since the thickness of
the sections (typically 40-50nm) is larger than the smallest features.

Our overall approach is to align pairs of images by maximizing correlation of portions of the images; this generates correspondence points which are used to place each image,
via an affine transform, into an approximate global coordinate system.  
Then each image is warped to better match the images above, below, and on the same layer.
We also generate a more detailed, non-affine transformation using a deformable mesh between each overlapping image pair.  
This transformation more accurately describes the relationship between each particular image pair, but is more difficult to extend to a global coordinate system.  
In our flow, each image is segmented independently.  
We use the pairwise transformation for linking structures across images, an analysis that needs only pairwise correspondences.

Our fundamental result is that we can align stacks consisting of 100s of thousands of images, with quality sufficient for neuron reconstruction.  
The computer time requirements are not great; alignment takes less than a day using a compute cluster of 500 computers. 
Successful alignment normally requires an image overlap of at least 100K pixels, though more is desirable.  The quality of the alignment depends on the defect density of the images.  For mosaics without folds or tears, 
alignment typically will give RMS residuals of less than 1 pixel within a section, 
resulting in a 'seamless' mosaic.  
Between sections, typical fits are about 3-4 pixels RMS with occasional larger errors where there are deformations that do not well fit our locally affine model.  
This alignment quality is sufficient to ensure that almost all proofreading difficulties are caused by legitimate ambiguities in the source image and not alignment problems.

The system can deal with many types of image defects automatically, but not all of them,
and which ones can be handled is not obvious in advance.  
Therefore the normal procedure is to try a fully automatic fit, then look at the metrics and residuals as discussed in section 11, {\it Debugging the Fit}.  
The images, and/or the alignment procedures and settings, can then be modified to create a better alignment. 
Typically the largest errors have causes that are fairly obvious and easily fixed, such as folds that begin or end in the middle of a section.  
However, there may also be places where the tissue has been sheared or warped in ways our relatively simple transformations cannot correct.  
These errors can be big (hundreds of pixels) but are relatively infrequent, hard to correct, and tend to occur on the edges of the mosaic or directly adjacent to folds.  
Since our reconstruction process included a manual component already, and since localized errors present little problem for a human proofreader, in general we ignored these errors and worked around them when needed.

One fundamental problem of alignment is finding correct but small overlaps without generating false matches.  
We would like to match small areas to minimize redundant imaging, and provide matches wherever possible to optimize alignment.  
However, in large regions of neuropil, there are many places that look similar, and can generate false matches.

An outline of this report is as follows.
In section 2, we discuss the normalization of images in the presence of defects.
Section 3 covers the automatic detection and analysis of folds.  Section 4 describes
finding the first approximate match, which is improved to the best affine match
in section 5.  
Section 6 then describes the deformable mesh between image pairs, and how it is computed, and section 7 describes how we try to distinguish true and false matches.  
Section 8 describes the mapping of the needed work onto a cluster of computers, then section 9 and 10 describe creating the global and detailed alignment respectively.
Section 11 describes how the user debugs the alignment process, and section 12 covers saving the transformations for later changes in the flow or alignment.  Section 13 discusses the results and 14 the conclusions.
\section{Read and normalize the images}
\begin{figure}
\begin{center}
\noindent
\includegraphics[height=7cm]{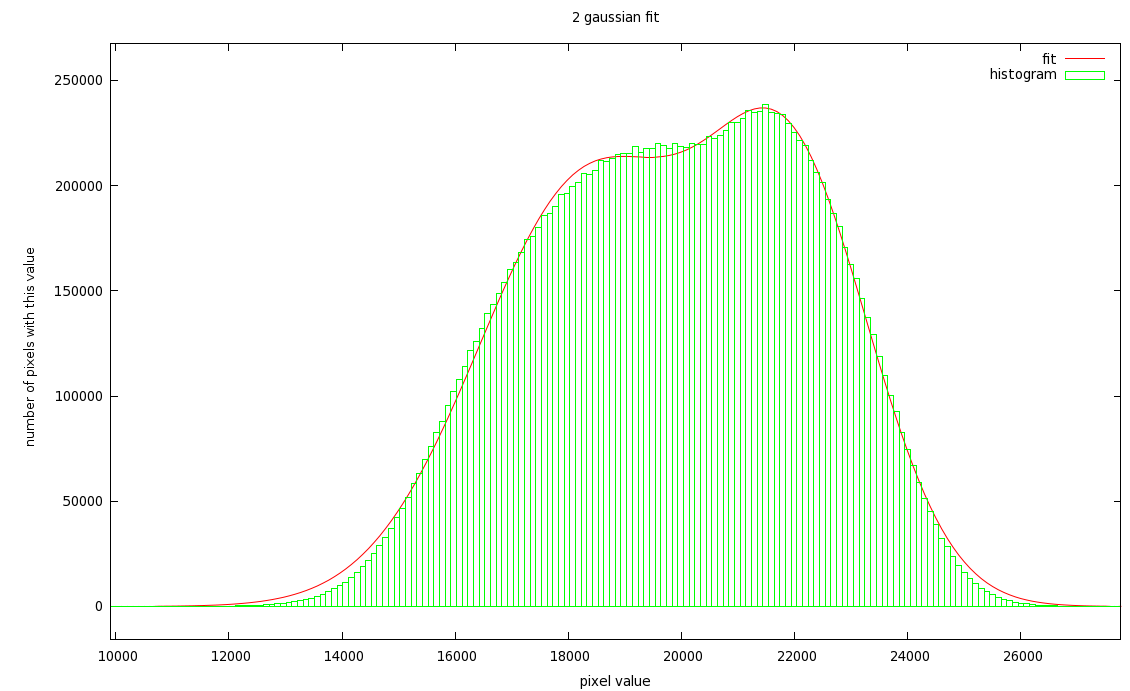}
\caption{Two gaussian fit to EM intensity histogram.}
\label{fig:Gaussians}
\end{center}
\end{figure}
Images from EM systems may be taken in a number of different formats, and are typically stored as 16 bit integer or 32 bit floating point values.  
Since the data sets are very large (terabyte data sets are not unusual), and many processing steps are memory intensive,  8 bit images are preferred.
This conversion loses very little information - if the original pixel collected $2^N$ electrons, 
then there is already a noise of $\sqrt(2^N) = 2^{N/2}$ bits just from Poisson noise in electron counts, resulting in a signal to noise ratio that is
at most $2^{N/2}$.  
So in the usual case where the microscope collected fewer than $2^{16}$ electrons per pixel, then the signal to noise is already less than $2^8$.  
So reducing the dynamic range to 8 bits hurts very little.  
We verified this by differencing several samples of original 16 bit images and 
appropriately scaled 8 bit versions.  
The only differences were featureless noise, as expected.

Normalizing the intensity of an EM image cannot be done by simply scaling the observed pixel value range to [0,255].  
This is because images may contain intensity outliers that are exceedingly dark (such as folds, see below) and/or unrealistically bright, such as tears  
or contamination.
Fortunately, these outliers are normally just a small fraction of the pixels in an image.
To take advantage of this, normalization uses the observation that the intensity histogram
of the neural tissue within an EM image is normally quite close to a sum of two gaussians,
as shown in Fig. \ref{fig:Gaussians}.
The normalization step creates a histogram of image intensity for each image and fits a two-gaussian curve to this histogram.
Then it takes the union of the $\pm 4 \sigma$ range of each of the two gaussians, and linearly converts this range to [0,255].
Since the information needed for neural reconstruction - membranes, vesicles, T-bars, and post-synaptic densities - resides in the darker portions 
of the image, our interactive tools also allow the user to bias the intensities so the darker tones are near the middle of the intensity range.
\section{Fold detection}
\begin{figure}
\begin{center}
\noindent
\includegraphics[height=7cm]{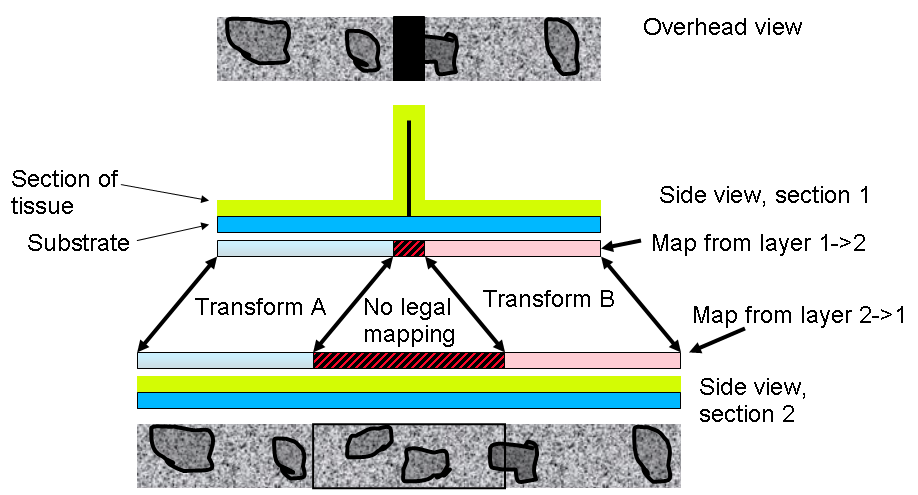}
\caption{Overview of folds and why piecewise transforms are needed}
\label{fig:Overview}
\end{center}
\end{figure}
Folds are locations where the tissue has been folded when placed upon the imaging grid -  see Fig. \ref{fig:Overview} for a cartoon example.  
Typically a fold is roughly 20 pixels wide in a 4 nm/pixel image, but the transformations needed to match the two sides of the fold
often differ by several hundred pixels.  There can also be folds in the substrate film, which have a similar appearance but do not cause
the alignment of the two sides to differ.
\begin{figure}
\begin{center}
\noindent
\includegraphics[height=7cm]{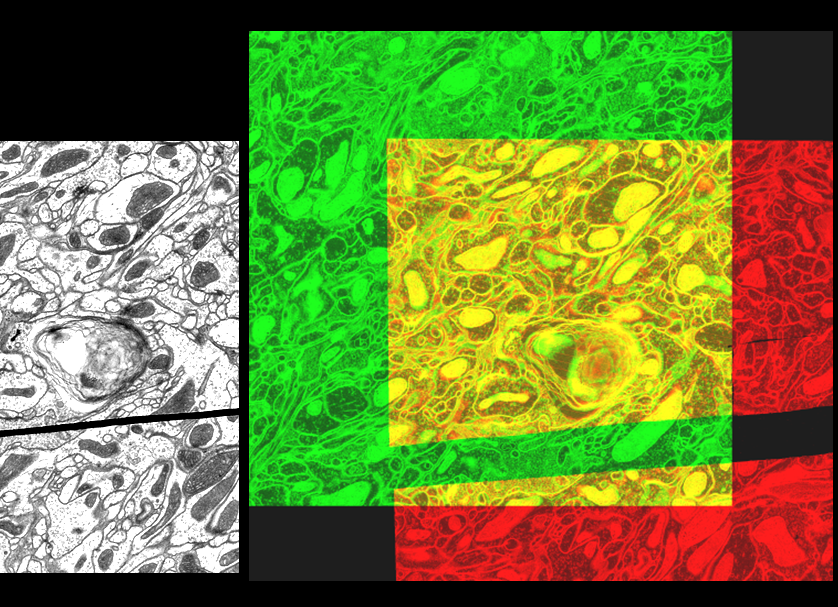}
\caption{The left side shows an image that includes a fold, which appears as a dark line.  The right side shows the same image, now in red, superimposed on an adjacent section, shown in green. The data on each side of the fold requires a different transformation for a correct match.}
\label{fig:Fold}
\end{center}
\end{figure}
In theory folds can be handled in at least two ways - there could be a continuous but very non-linear transform across the fold, 
or the image can be divided at the fold and the two sides of the fold treated independently.  
Since treating each side of the fold independently is needed in any case during segmentation and linkage,
we treat each portion independently in all phases.

The process starts by finding the folds, or locations where the images should be divided for the purpose of alignment.  
The fold is much darker than the regular image, as shown in Figure 3.
Therefore we start with the whole image, then remove pixels that are darker than a threshold, typically taken as 3 standard deviations darker than average.  
The remaining pixels are grouped into connected components (called patches), and small patches are dropped.  
Note that patches have very irregular outlines, and may contain holes, typically caused by contamination which can also create very dark spots on images.

The result of fold detection must be used consistently in several places.  
It must be used in alignment, for deciding the patches that will be transformed separately, in segmentation (segmentation must stop at each fold), 
and in linkage (linkage is the connection between sections - here a fold represents missing data, and linkage cannot continue in the normal manner through a fold).  
Since all of these steps need the same data, and since it is critical they all be consistent, the {\it fold mask} is computed once, then stored.  
The fold mask is a compressed image, with fold pixels given a value of 0, and all pixels of each patch containing the same integer value.
Thus all the pixels of the first patch have value 1, all the pixels of the second patch value 2, and so on.  
This map image compresses very well, and typically only requires space of about 1\% of the original image.  
Thus the overhead of saving, rather than computing, the fold mask is slight, and has the advantage of consistency across all applications.  
This is a non-trivial advantage since image processing applications are commonly written in a variety of languages (our flow, for example, uses a combination of MATLAB, C++, and Python).
\section{Finding rough matches}
After fold detection, the next step is computing matches between images.   The task of matching images has very different requirements
depending on whether the two images are from the same or different sections.

Within the same section, the angle between the images is small, and the transformation will be mostly translation.  
Differences in scale will be small, and there will be little skew in the transformations.
There is no need to consider folds since a fold, if present, will be present in both images.  
The final correlation will be very high (typically 0.8) since exactly the same tissue is imaged within the overlap area.

The situation is quite different when attempting a match between different sections.  
Each patch must be matched independently, since otherwise folds make close matches impossible.
Very little is known about the orientation and overlap of the images.  
There may be scale changes (a few percent is not uncommon), and sometimes there is
considerable skew between the axes of two consecutive sections.  
The final correlation will not be nearly as good, since only the larger features match - 0.35 is typical, rather than 0.8.

Therefore the matching process must be prepared to search through both angle and translation to find matches.  
It starts by creating reduced size images (roughly 256x256), which we call {\it thumbnails}.
For images of this size, the angular displacement that will change the correlation appreciably is about 1 part in 200, 
so this is the quantum of angle while searching for the initial match.
The program searches for the best alignment with a metric of  normalized cross correlation (using FFTs for the correlation and cumulative tables for the normalization).\cite{lewisfast}

One difficult problem is false matches during the thumbnail matching phase.  This can happen in at least two common circumstances.  
One is that there is a large blank area in each of the two images (perhaps from resin at the edge of the sample).  
This will naturally match very well to any large blank areas in any other image.  
The second happens if small overlaps are allowed - this can create false matches from similar-looking large features - for example, many mitochondria look alike.

Looking at many false and true matches, the true matches have a relatively compact and symmetrical peak in the correlation map.  False matches tend to have a more distributed and less symmetric structure.  Therefore we process the correlation map to bring out the structure types most likely to be true matches.



Finding the best angle between two thumbnails benefits in both accuracy and efficiency from the principle that all of the thumbnails composing 
the mosaic for layer $j$ have the same global angle $A_{jj}$, with variation typically less than 1 degree, depending on the accuracy of the stage.
Likewise, mosaics from differing layers $j$ and $k$ have a common global angle $A_{jk}$ between them, 
though the variation may be a degree or more due to differential warping of the slices. 
To find $A_{jk}$ (or $A_{jj}$) we use an adaptive process. 
For the first few thumbnail pairs for layers $j$ and $k$, $A_{jk}$ is as yet unknown so we use a very wide sweep of test angles $A_{jk}$ looking for the best correlation score. 
Each winning $A_{jk}$ is tabulated, along with layer labels $j$ and $k$. 
After four or more $A_{jk}$ have been collected, subsequent thumbnail pairs use a much tighter angle sweep about the median $A_{jk}$. 
The angle tables are retained and are very useful for debugging errant matches.

Once we find one or more approximate matches, we tune up the best match.  We do this by repeatedly trying small skew and scale changes, and keeping them where they improve the
match.  We stop when we find no more improvements.

\section{First Affine Match}
At this point, the next step is to return to working on the larger images (sampled down to 2k by 2k, if they are originally bigger).  
We work in two stages; first trying to find a good affine match, and then (if the patch is big enough), 
decomposing the patch into triangles, and allowing a different transform in each triangle.

In the first phase, we try to find a single affine transformation that gives the best overall match.  
This forms the starting point for the later deformable mesh calculations.
Also we use the affine match to update the lists of which pixels map between the two images, since this can be quite a bit off using the original transforms computed from the low resolution images.

For code simplicity, the affine match is found with the general mesh algorithm, but with just three control points defining one triangle.
We assign all points in the patch to this one triangle, then move the vertices of that triangle (by gradient ascent) to get the best match.
Since three points define an affine transformation, mathematically this is exactly the same as seeking the best affine transform directly. 

\subsection{Creating triangles}
After the affine match, we re-calculate the points that map from one patch to the other.  
If the number of points is high enough, we sub-divide the patch into smaller triangles.  
Typically we find that about 100K pixels per triangle is best.  
If the triangles are smaller, some regions will be nearly featureless and lead to convergence problems.  
If they are much bigger, then the differences between sections can prevent an accurate match.

Two techniques are supported to create a mesh of triangles.  
One tries to follow the irregular outline of each patch, while the other creates a uniform mesh of triangles.

\begin{figure}
\begin{center}
\noindent
\includegraphics[height=6cm]{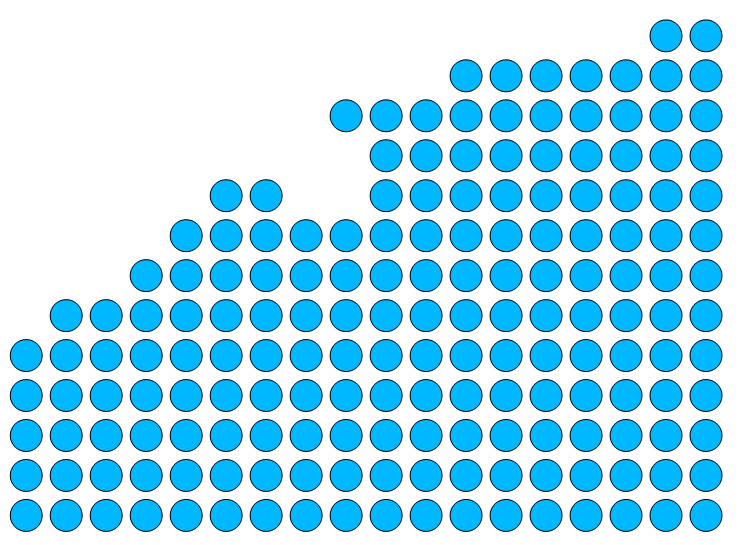}
\caption{Each patch may have an irregular outline}
\label{fig:Irregular}
\end{center}
\end{figure}

\begin{figure}
\begin{center}
\noindent
\includegraphics[height=6cm]{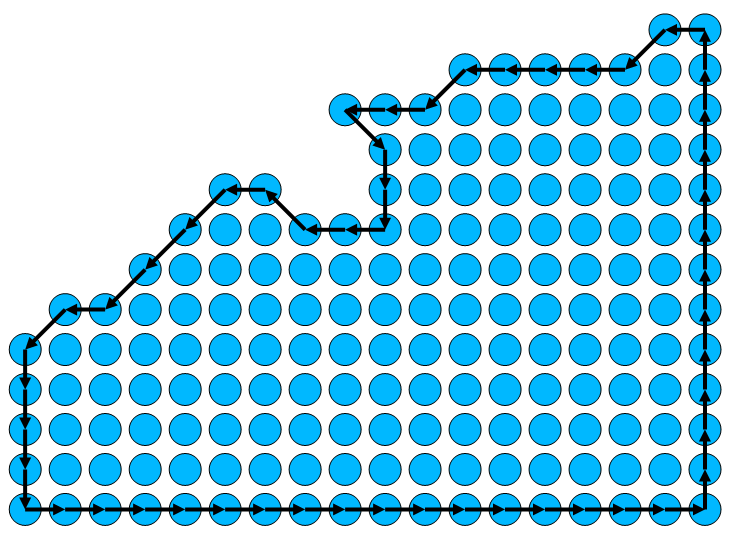}
\caption{Work all around the connected region, one pixel at a time}
\label{fig:DetailedBoundary}
\end{center}
\end{figure}

\begin{figure}
\begin{center}
\noindent
\includegraphics[height=6cm]{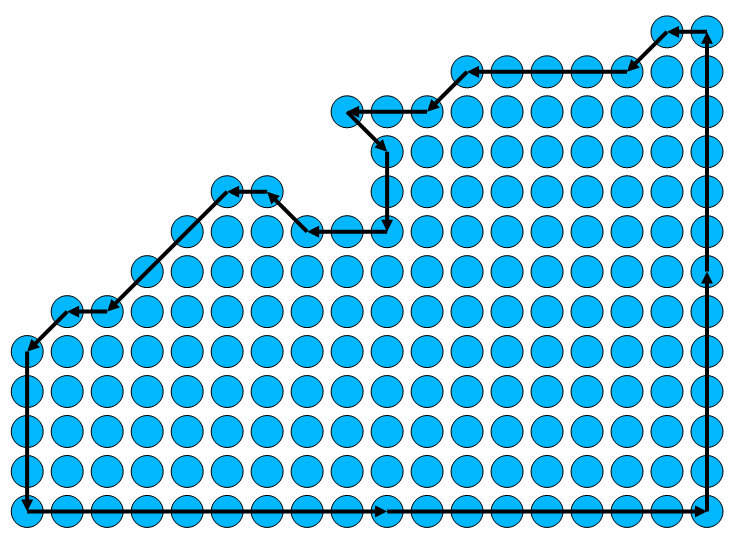}
\caption{Combine edges that point in the same direction, then split edges that are too long.}
\label{fig:CombineEdges}
\end{center}
\end{figure}

The contour hugging algorithm starts by creating a pixel map of the pixels that map according to the initial affine transform.  
Because of folds and other artifacts, these regions may quite irregular - see Fig. \ref{fig:Irregular}.
The algorithm works around the map counter clockwise, starting from an initial pixel.  
This generates an outline composed of short vectors (8 directions possible at each vector), as in Fig. \ref{fig:DetailedBoundary}.  
Then it combines like vectors together and splits vectors that are too long, as shown in Fig. \ref{fig:CombineEdges}.  
Then we find the best approximate outline that can be constructed with edges of restricted (min,max) length.  
The cost of each vector is the sum of the distance of all intervening points from the vector.  
Using an A* search,\cite{hart1968formal} we find the lowest cost set of circumscribing vectors of legal lengths.

At this point we have a polygon of vectors.  Next, we add interior points if needed (if there are interior points too far from all existing vertices).  
Then we triangulate by taking an arbitrary edge, finding the best interior triangle from that edge, then removing this triangle from the original polygon.  
We repeat until nothing is left or all remaining triangles are too small.  
Finally, we assign each point to the nearest triangle, as shown in Fig. \ref{fig:IntoTriangles}.  
Because the outlines of the patches are irregular, there can be both points outside a triangle that are assigned to it, and interior points which are not assigned.

The uniform mesh is simpler to create, with only the suggested size of the triangles to be specified.  
In our experience, this simpler approach has fewer corner cases and is more reliable.
The intersection region between two rectangular images is typically an irregular polygon. 
To approximately tile this region with triangles we construct the rectangular bounding box that circumscribes the polygon. 
The box is then subdivided into a simple grid of equal sized right triangles. 
Finally, we compute the actual intersection area of each triangle with the 
original polygon and if that is below 75\% of the triangle's area, 
the triangle is removed from the mesh. 
The resulting mesh need not conform exactly to the intersection region. 
All that is required is that any point in the region be 
deterministically associated with some triangle so that the corresponding 
transform (mapping) can be applied for that point.
\begin{figure}
\begin{center}
\noindent
\includegraphics[height=6cm]{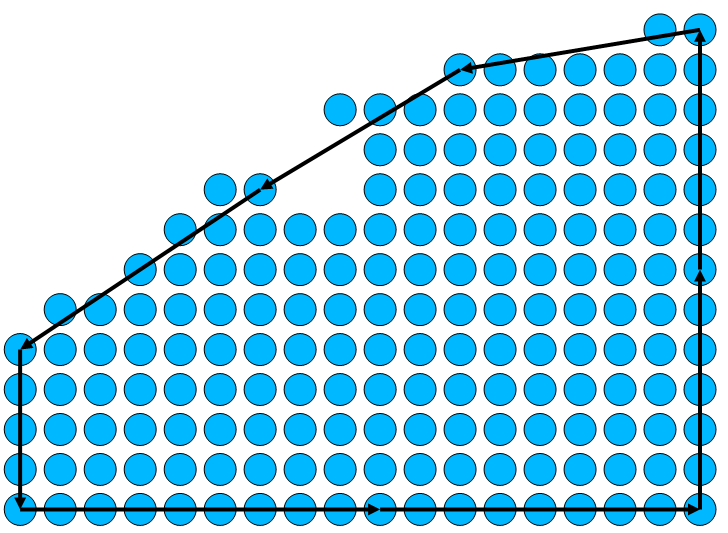}
\caption{Find the best approximation to an outline that uses only edges that are not too short, and not too long.}
\label{fig:ApproxOutline}
\end{center}
\end{figure}
\begin{figure}
\begin{center}
\noindent
\includegraphics[height=6cm]{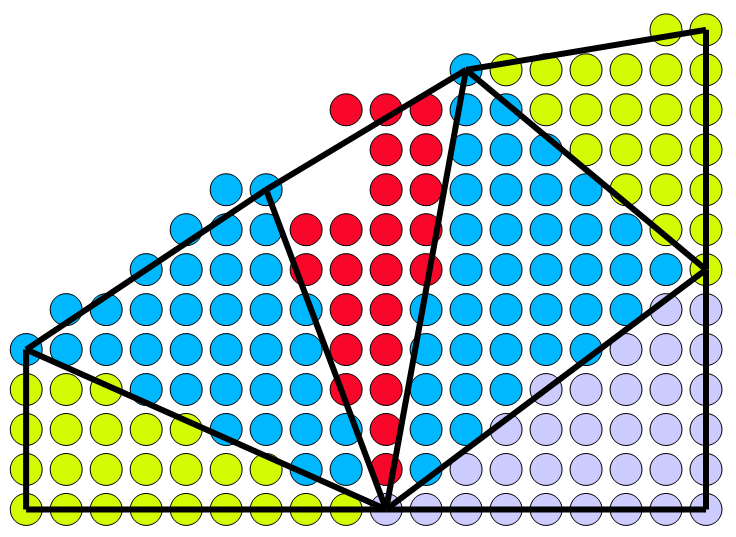}
\caption{Divide into triangles, and assign each pixel to the nearest triangle.}
\label{fig:IntoTriangles}
\end{center}
\end{figure}
\section{Searching for better correlation}
The next step is to move the control points to get a better correlation via a simple gradient ascent algorithm.
This is done in exactly the same way, whether the control points form one triangle, or many.  
First, each pixel is associated with one triangle (though it is not necessarily inside, see Fig. \ref{fig:IntoTriangles}).
Then the XY coordinates of the pixel are described as a linear combination of the 3 vertices (also known as control points) of the associated triangle.  
This is always possible since the vertices of a triangle cannot be colinear.
Then the detailed correlation is computed by summing over all points associated with a patch.  
Using the current linear combination of control points, we find for each point the current XY position in the other image.  
(This will usually not be an integer pixel coordinate).
Then we interpolate bi-linearly in the target image to get the value of the target, multiply by the source pixel value, and add to the correlation sum.
At the same time, since both the point location as a function of the control points, and the local derivative in the target, are linear, 
we also can accumulate the derivative to the correlation, describing how the correlation changes if any of the control points are moved.  
The matcher then uses this derivative information in a simple uphill search.
Since the exact derivatives are computed each time the correlation is updated, there is no need for a sophisticated algorithm that minimizes calculation of the derivative.
Instead the program starts with a total move of 10 pixels length in 2N dimensions (where N is the number of control points),  
and steps in the uphill direction.  If the correlation is better, the move is accepted;
if not, the step size is cut in half.  The process stops when the step size becomes less than a threshold, normally set to about a tenth of a pixel.

\section{Quality checks}
Once a proposed match if found, it is tested in several ways to see if the match is legitimate.
The first test is the value of correlation found.  This is all that is needed for in-section matches, where the correlation is very high, particularly after image deformation.
This test is much more ambiguous on inter-section matches, where intermediate values (near 0.3) are common.

The next test looks at the Fourier transform of the image differences, after matching.  The intuition is that for a correct match, the power in the difference image
should be concentrated in the high frequencies.  The large features, such as mitochondria, should match well, resulting in little power in the low frequencies.  This
test calculates the cumulative power in the difference image, from low to high spatial frequencies, and computes the half-power frequency (half of the power is below, and half above).

The next test is the number of ``yellow'' pixels. This name derives from our diagnostic images, where one image is drawn in green, then the other (warped to match)
is drawn in red on top of the original image, as shown in Fig. 3.  
For a good match, we expect most pixels to be yellow, since they will have roughly the same amount of red and green.

A final test is the value of the earth mover metric on the difference images.  The Earth mover metric is the minimum number of quantum pixel moves needed to convert
one image into another.  Each quantum move moves a single pixel delta of 1 by one pixel.  (If you think of the difference image as representing the height of a landscape,
then this is the total amount of dirt-moving that is needed to flatten this difference image).  
This metric is difficult to calculate exactly\cite{peleearth}, but can be approximated
by using a wavelet decomposition, then weighting the amplitudes by the wavelengths.

Each of the above metrics is compared to a user specified acceptance threshold.
We normally set the metrics so no false matches are accepted for the entire data set.  
This is because one false match will destroy the solution quality of a least squares fit, whereas missing correspondences are (largely) compensated for, since each tile
typically has correspondences with 12 others.
\section{Software Engineering}
The alignment must work as automatically as practical on large data sets (for example, one medulla data set has 9x9x1900, or about 153K images).  
Furthermore, each image must be aligned with several others (at least 4 on the same layer, 4 each on the layers above and below).  
Overall millions of pairs of images must be aligned.  
Therefore a significant fraction of this work has been devoted to making this process
amenable to parallel computation, so it can be completed in a practical time scale.

The process starts with a rough global alignment, which can be obtained from other software such as TrakEM2,\cite{cardona2006trakem2}
or from the microscope control software Leginon.\cite{potter1999leginon}
The rough alignment is used primarily to determine which tiles may potentially overlap, to avoid the need to check all tiles against each other.  
In our case in particular, the initial alignment was performed in 20 section blocks,
due to practical limits in TrakEM as of 2008.  
A crude alignment between the adjoining faces of adjacent blocks was obtained by matching the center 4 images of each face against the center 4 images of the adjoining face.  
This alignment was then extrapolated to the other images of the adjoining faces.

The next step created one {\it make} file\footnote[1]{A {\it make file} is a text file that describes operations to be performed on files, and includes explicit
information describing how each output or intermediate file depends upon its input files.  
The operation(s) are performed if either the desired output file(s) do not exist, or are older than at least one of the files on which they depend.}
per layer (for in-section alignment) and two make files per adjacent layer pair, one aligning from layer up, and another from the upper layer down.  
In the case of the medulla this yields about 3800 scripts, 
each of which can be submitted in parallel to a compute cluster.  
Each script compares each image in a given layer to all likely overlapping images - North, South, East and West images for a single layer computation, and the (typically four) likely overlapping images for an adjacent layer computation, as computed from the crude alignment.
Each script takes an hour or two to run, and our cluster contains 500 computers, each with 8 cores.  This reduces the elapsed time from roughly a year (on a single machine) to half a day.

Constructing scripts as `make' files has two benefits.  The first is the traditional
benefit of efficiency - if one image is modified (most often to make a fold more explicit), then only the
results that depend on that image will be re-computed.  
This is extremely helpful when debugging alignments.

The second benefit is that make scripts can be given additional
parameters that are passed to each program's command line.  
This allows easy handling of special cases.  
On data sets of this size, there are typically exceptions from the regular rules.  
For example, in a stack of images of the lamina (see section 14), there is a subset of the images have a scale factor that is 10\% different from all the others.  
This is enough to prevent the program from finding an acceptable match.  
There is a command line option to take this into account, by scaling one image as part of the correlation process, 
but if applied to layers where the scale is the same it will cause missed matches.
Therefore it is important to use the option only in the cases where the scales differ significantly.    
Similarly, there are cases where certain layers have skew between the X and Y axis, low contrast in certain portions, and so on.  
In general these can be taken care of by use of specific options, but if these are applied to all runs then too many false matches will result.  
So we wish to apply different parameters to just a subset of all alignment runs, and need a reliable and reproducible way of organizing exceptions.  

This can be taken care of in the program that creates the make files, 
or by modifying the individual make files themselves.  When making the make files, 
the user can specify in a file any number of triples consisting of 
(layer, layer, extra arguments).  
These are applied whenever the two layers mentioned are processed.  
In the example above, layers 191-200 had the smaller scale.  
Therefore the commands were (190,191,``-SCALE=0.9'') and (200,201,``-SCALE=1.1'').  
The scripts that are generated to align these layers have the extra arguments added, and can be run (or re-run) automatically.  
It is also possible to run exceptional cases by hand, or to add them to the make files
directly, but adding the exceptions to the script generation both documents what 
needed to be done, and allows it to be repeated if necessary.

\section{Global alignment}
We generate aligned data in two passes.  In the first pass, we compute one affine transform per patch.
Although not the best possible match, this alignment is usable by many different programs such as TrakEM, Raveler\cite{Raveler}, and so on.
To compute this global affine we take matching points, 
derived from our deformable mesh fit, then compute transform values to make them match as well as possible via least squares fit.  
In addition we need a few more constraints.  If we express the transformation for each patch as
$$ \left[ \begin{array}{c} x\prime \\ y\prime \end{array}\right] = 
 \left[ \begin{array}{ccc} a&b&c\\d&e&f\end{array}\right] 
 \left[ \begin{array}{c}x\\y\\1\end{array}\right]$$
then we add the following constraints:
\begin{itemize}
\item One image is defined to not change, so, $a = e = 1, b = d = 0$.
\item We prefer transforms that look like rotations with scaling. Such transformations have $a = e$ and $b = -d$.  Therefore we add desired relationships to the least square fit: $W_s(a - e) = 0, W_s(b + d) = 0$, where $W_s$ is the strength of the desire to keep the images square.
\end{itemize} 
We do not constrain the absolute scale at this step.  This is because the constraint we really want is $a*e-b*d = 1$, which is not linear.  We solve this by a two pass approach.  In the first pass we ignore this constraint. 
This will always introduce some scaling error if 3 or more layers are 
involved (see Appendix).  
The error introduced is small if only a few layers are aligned,
but grows as the number of layers squared (to first order), and starts to become very significant at about 100 layers.  
This initial solution, however, provides us with the correct relative angles between the sections.  
Once the angles are known, scaling can now be expressed as a set of linear constraints. 
We add these constraints and re-solve, as described in the Appendix.

Since the number of images may be large, there may be many equations and unknowns in this process.  
For example, a medulla imaging and reconstruction involved 81 images per section, and about 1900 sections, for a total of 153K images.  
The least squares fit for all affine parameters then had more than 918K variables and (in our case) about 14M constraints.  
Fortunately the constraints are very sparse, so traditional sparse matrix techniques can be used to solve these problems.  
Our code converts the constraints to the normal equations, then solves with the sparse matrix package SUPERLU.\cite{li2005overview}
It takes about 10 minutes on a (circa 2010) workstation to solve a problem of this size.

This first stage of the process results in an affine transformation for every patch.  
This can be viewed in TrakEM or used to construct images for our in-house proofreading tool `Raveler'.  
However, with the limited flexibility allowed by a single affine per patch, the alignment has typical errors of 9-10 pixels RMS, particularly between sections.  
This is good enough for most manual proofreading, as the human eye is very forgiving of alignment errors, 
but it can cause problems for both humans and machines 
where fine processes cross tile or section boundaries.  
This is the motivation for the next step, detailed alignment.
\section{Detailed alignment}
\begin{figure}
\begin{center}
\noindent
\includegraphics[height=6cm]{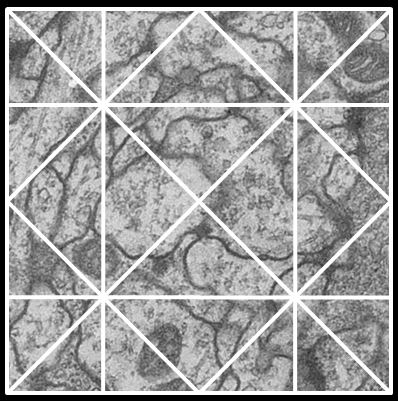}
\caption{The finer mesh used to warp each image into global coordinates consists of 24 triangles with 21 control points.  The smaller triangles are concentrated in the corners, where the distortions are more severe.}
\label{fig:Detailed}
\end{center}
\end{figure}
In this step, each patch is allowed to deform further.  
Again we divide each patch into triangles, different from the triangles of section 7 (Those were intended for optimum pair-wise matching, these for tying multiple images into a common framework.)  Since image deformations tend to be worse at the edges and corners of the image, we allow a deformation as shown in Fig. \ref{fig:Detailed}.
This divides each image up into 24 triangles defined by 21 control points, or 42 degrees of freedom in all.
Starting with the overlap as defined by the affine alignment, we find local correspondence points in all overlapping images, again by normalized cross correlation.
Each correspondence point between two patches is expressed as a linear combination of the three vertices of its containing triangle in each patch.  
This sets up two linear equations (one X, one Y) in 6 variables for each correspondence point.  
For regularization, we add an additional constraint for each control point that constrains it 
to be close to where it would appear under the affine transformation computed above.
We then solve for the location of the control points that minimizes the mismatch of the correspondence points, using the same least squares solver described above.

We used this technique independently on each layer to 'heal' the boundary between images.  This worked quite well, yielding a 'seamless' stack where the edges between each image were not visible.  It should be possible to extend this to cross-layer alignment as well, though we did not try this.
Since each image now has 42 parameters instead of 6, and correspondingly more constraints, it may no longer be possible to solve the whole stack in one run.  
In this case the technique one solution might be to solve as many layers as practical, say [0..N].  
Then layer N-10 is fixed, and serves as the bottom layer of a run that solves for [N-10..2N]. The next run runs from [2N-10..3N], and so on.  
This will generate a slightly less good solution for layer N-10, since it can 'see' only 10 layers in one direction, 
but it will be very close to an (impractical) full solution since the effects of layers more than 10 sections away is minimal.

Using the computed deformations, the program creates the final images used for
proofreading.  
In areas of image overlap, the program has a choice of which data to use.  
Since deformations tend to be largest in the corners of images, it uses (for each output pixel) data from the image whose center is closest.  
When done drawing, the program records all transformations that were used, 
as discussed further in section 12.
It generates output files, in at least three forms - a large flat image for each section, a set of tiled images for our proofreading
program `Raveler', and a file of transformations for TrakEM.   
It also transforms auxiliary files, such as segmentations and annotations, 
that were computed or stored on the image tiles and must be mapped into global space in synchronization with the grayscale images.

\section{Debugging the Fit}
It's not always easy to debug a global fit with many variables.  
As an example, consider a small triangular patch in the corner of a tile, perhaps isolated by a fold.  
Such a patch might have only two correspondence points with neighboring tiles.
This makes the fitting procedure ambiguous, with two solutions reflected across the line between the two points.  
Only one of these is correct, but both solutions are mathematically perfect and hence do not show up in the quality of the fit. 
To avoid this problem, we can demand that each patch have at least 3 points, but then this region shows up as a hole.

Typically, the user starts by ensuring each section can be fit into a mosaic by itself, looking at the number of matched points and the error metrics from the least-squares step.  
Once all single sections work, then all layer pairs, then layer triples, are tried next. 
If these work then larger blocks are usually OK.

If any of these alignments fail, the user must figure out why and fix the problem.  To help with this, the alignment programs produce an array of debugging outputs.
These include text reports then show the worst mis-matches, sorted by layers, by layer pairs, and by the particular images involved.  They also include
various plot files which can be viewed with 'gnuplot' or other plotting utilities.  These plots show a schematic representation of the fit, with outlines of tiles
and correspondence points shown in various colors.

The most common solution to alignment problems involved extending a fold to the end of a section.  Next would be changing some parameters of the alignment scripts to accommodate image changes such as differences in scale or focus.  The least used fix was manually specifying alignment points; this was used when entire sections or portions of sections were missing.  The need for this could potentially be eliminated by comparison of non-adjacent sections, but since this was a rare problem we did not attempt it.

The manual effort required to fix the alignment is difficult to specify exactly, since it was intertwined with the software development.  
Often the choice, given an alignment discrepancy, was to fix the alignment directly, or improve the software to deal with it.  
The sum of the time needed for developing and debugging the software and fixing the alignment was about two person-years.  
As a crude estimate, it might take one to two months to align a similar data set if the software remained unchanged.  
We have not tried this explicitly since our next experiment used FIB-SEM technology, which has its own and very different alignment challenges.

\section{Save the Transforms}
It is critically important that all mapping transformations be saved along with the generated images.
Often it is the case where a better (or at least different) alignment is needed, 
but some manual work has already been done on the stack.
In this case the corrected results must be transferred to a new alignment.  
In our flow,  this process works backwards from the corrected data, and maps the corrected annotations back to the tiles from which they came.  
These tiles can then be assembled into a new alignment.  

To enable this mapping back into tiles, and then forward into new alignments, the programs that compute the alignments also produce `mapping' files, which are image and text files that tell, for each pixel in the output stack which image that pixel came from, 
and what transformation was used to map it from the tile to the final stack.  
Since many pixels share the same transformation, again these files compress very well and take up negligible space compared to the grayscale images.

\section{Results}

We have applied this software to two neural reconstruction projects.  
The first was of the lamina of the fruit fly {\it Drosophila melanogaster}.\cite{Marta}  
Here we aligned a 6x6x700 array of 2Kx2K TEM images (6.2nm per pixel), with very small overlap at the edges (only 1\% in some cases).  
The second project was reconstruction of the medulla of the fruit fly; it involved a 9x9x1900 array of 4Kx4K TEM images at 3.1 nm/pixel.
However, these images had more overlap (about 5\%)\cite{Shinya}
The 5\% overlap was enough that there were no false matches, or missed matches, in the full data set.  In the 1\% overlap case, we could not find any settings that
found all true matches without any false ones, and a few (about 10) manual parameter settings were required to complete the in-section overlaps.

When there are no folds or tears, the resulting RMS errors in the affine fit stage were typically about 9 pixels, with the largest contributions from lens distortion and sample shrinkage during imaging.
After the final deformation, the process can normally result in sub-pixel matches within a single layer, and roughly 2-3 pixel matches between layers.
Since the sections are typically 40 nm thick, with 3-4 nm pixels, a process at a 45 degree angle to the cutting plane would move 10-15 pixels 
between adjacent sections.  Therefore it is not at all surprising that the fit between sections is not as good as the fit within a single section.
This alignment appears good enough for our purposes - the largest remaining source of errors and ambiguities is not inadequate registration, it is the 
difference between adjacent sections due to their physical thickness.

However, in both cases the worst fitting errors were on the order of 500 pixels, too big for the deformation of images to correct.  
Investigation showed the largest of these errors were caused by folds that began or ended within a section.  
The solution chosen was to run the fit, find these errors, then manually continue the folds to the edge of the image.
In the stack of 153K images, 49 images on 31 layers were modified, reducing the worst-case error to 100 to 200 pixels.  
Examination of some of the remaining errors showed that many appeared to be non-affine distortion of the tissue caused by the cutting process that did not rise to the 
level of an explicit fold or tear.  Since the proofreading process could cope with occasional mis-alignment of this magnitude, we did not investigate further.

There is no obvious impediment to working on much larger data sizes, but these data sets were already bigger than we could proofread in a practical amount of time.
Therefore larger data sets have not yet been tried.

As an experiment, the same programs have been used to align other image types as well.  
Mosaics of optical and scanning electron microscope images can be handled with only
parameter changes, though further optimization would be needed were these to be used in production.
\section{Conclusions}
Stacks of hundreds of thousands of electron microscope images can be aligned to the accuracy needed for neural reconstruction despite the
presence of the most common image defects.  A 5\% image overlap is enough to ensure automatic alignment within each section.
Contrast and intensity differences, image distortion due to sample shrinkage, and artifacts due to contamination
are handled automatically.  
Folds and tears in the tissue are handled largely automatically, but require manual intervention in a small percentage of
cases.  The process parallelizes well, and can be performed in about a day on a 500 computer cluster for the data sizes considered here.
The same alignment process works on other imaging modalities as well.

\section{Acknowledgments}
We gratefully acknowledge the help of our long suffering users, who painfully generated biological results by using our software while it was still
under development.  These include Marta Rivera-Alba, Shinya Takemura, and Ian Meinertzhagen.  

%
\bibliographystyle{abbrv}
\bibliography{align}
%
%
\section*{Appendix - why least squares fit introduces scaling problems}
\begin{figure}
\begin{center}
\noindent
\includegraphics[height=7cm]{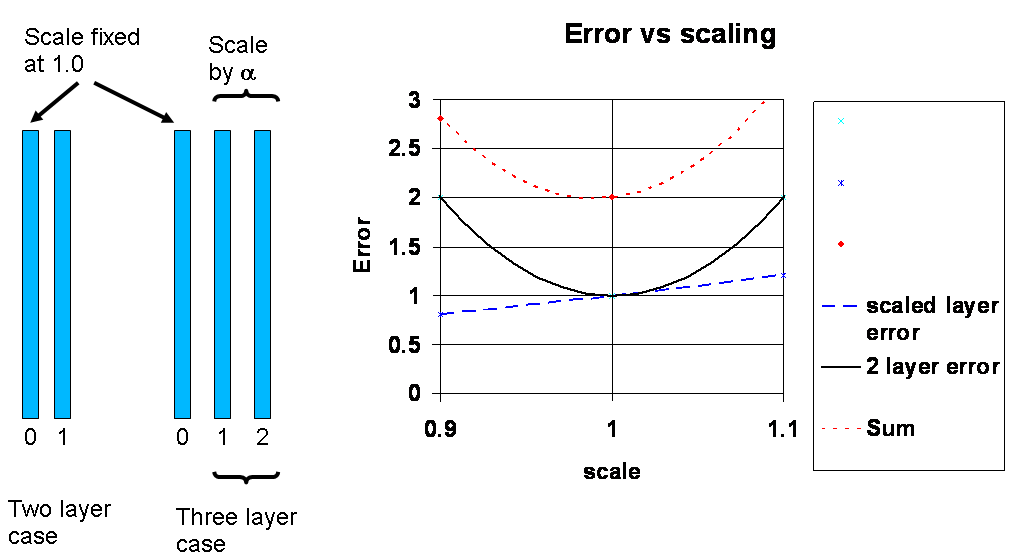}
\caption{The scaling problem - why pure least-squares does not work.}
\label{fig:Shrink}
\end{center}
\end{figure}
Let $C$ be the number of connections, $e$ the mean square error per connection, and $E$ the total mean square error per layer pair, so $E = C e$.
Now imagine a triple of layers, called 0, 1 and 2.  
Fix the scale of the first layer at 1.0.  
Now change the scale of the second 2 layers to $\alpha$. Now the 1-2 errors scale as $\alpha^2$.
Assume an image $N$ by $N$ pixels, with the origin in the center.  Assume the errors between sections are independent and normally distributed.

The added error in $x$ is $(1-\alpha)*x$, and in $y$ it is $(1-\alpha)*y$.  On the assumptions the errors are independent and add in quadrature, we get the total 0-1 error as:
$$E + \sum_C{(1-\alpha)^2 x^2} + \sum_C{(1-\alpha)^2 y^2} $$
which is also
$$E + (1-\alpha)^2 (\sum_C{x^2} + \sum_C{y^2})$$
Now assuming that x values are uniformly distributed between $-N/2$ and $N/2$. we get the average value of $x^2$ will be
$$1/N \int_{-N/2}^{N/2}{x^2 dx} = \frac{1}{12} N^2$$
The Y values contribute equally, so the total 0-1 error is
$$E + (1-\alpha)^2 C N^2/6$$

Fig \ref{fig:Shrink} shows these two components, and shows the fundamental problem - the total error is the sum of two components.  
One has a minimum at $\alpha=1$, the other increases quadratically from 0 and always has a non-zero slope at $\alpha=1$.  
Therefore the sum will always have a minimum at some $\alpha<1$.

First we express the total 0-1-2 error as
$$E + (1-\alpha)^2 C \frac{N^2}{6} + \alpha^2 E$$
Taking the derivative with respect to $\alpha$, and setting it to 0, we get
$$\frac{C N^2}{6} 2 (\alpha-1) + 2 \alpha E = 0$$
And solving gives 
$$ \alpha = \frac{1}{1+\frac{6 E}{C N^2}} = \frac{1}{1+\frac{6 e}{N^2}}$$
The final result is independent of $C$ since it affects both errors equally.  
Plugging in typical values ($e=9, N=4000$) we see that for three layers the effect is very small, about $3.3\times 10^{-6}$.   
With $N=4000$ this only results in worst case error (at the edge of the image) of 0.013 pixel.  This is negligible in practical terms.

However, as the number of layers increases the effect becomes bigger, since the penalty of a scale factor change from layers 0-1 is now balanced by
a gain in all the layers 1-N.  Call the original error computed above $\alpha_1$.  
Then repeating the calculation shows the next error, $\alpha_2$, will be roughly $1+\alpha_1$ times bigger, or very nearly twice as big. 
The next layer will have error $\alpha_3 \approx 1 + \alpha_2 (1 + \alpha_1)$, or roughly 3 times as big.  If we let the weighted sum of all the
layers 1-N by $R$ times the error in the original case, and re-do the calculation above, we find that
$$ \alpha = \frac{1}{1+\frac{6 e R}{N^2}}$$

Now we can solve for the scale factor changes.  In general, the equations are complex, but we can derive simple approximations for both relatively
small (a few hundred) and large (more than a few thousand) layers.

First, for the small number of layer case we
assume that $\alpha$ is close to 1 (since we will terminate the process if any of the ${\alpha}$s stray far from 1.0), then we find  $\alpha_N \approx N \alpha_1$.
Then the total scale change from layer 0 to layer $N$ will be roughly
$$\prod_{i=1}^{N-1} \alpha_i \approx 1-\sum_{i=1}^{N-1}(1-\alpha_i) \approx 1-\sum_{i=1}^{N-1} i (1-\alpha_1) \approx 1-(1-\alpha_1) \frac{(N-1)(N-2)}{2}$$
$$\approx 1-(1-\alpha_1) \frac{N^2}{2}$$
Thus for about 400 layers, the effect will be about 80,000 times bigger,
or about 25\% in linear scale for the assumptions above.  This is quite noticeable.  For 1000 layers, it's more than a factor of 2.

In the limit of a very large number of layers, the shrink per layer approaches a constant.  
This happens since the total effect of all the layers to the right does not grow indefinitely, but is instead the sum of a series of decreasing values.  
In the limit, where the shrink per layer approaches a constant, this is the sum of a geometric series.  
Thus if the limiting shrink per layer is $\alpha$, then
$$R = 1/(1-\alpha)$$
which leads to the equation
$$\alpha = \frac{1}{1+\frac{6 e R}{N^2}\frac{1}{1-\alpha}}$$
Solving for $\alpha$, we find
$$\alpha = 1 + \frac{3 e}{N^2} - \sqrt{\frac{6 e}{N^2} + \frac{9 e^2}{N^4}} $$

Of course the model is simplified, the error is computed only to first order, the in-section alignment is neglected, and so on.  
But the general result is true - using least-squares alone results in sizable drifts in image scale, and they start to become significant at about 100 sections.
\subsection{Experiments with artificial data sets}
We back up this hypothesis as follows:
\begin{itemize}
\item Create synthetic data sets.  We use a program that creates tiles and correspondence points at random.  The points are uniformly distributed, errors in the points are normally distributed.  For the examples below, we use N=4000, 20 correspondence points per layer, and an E of 18 (normally distributed errs with a standard deviation of 3 in both X and Y.
\item Solve using least squares.  We construct the normal equations in the usual way (preserving sparseness) and then add the following constraints.  
$a = e$ and $b = -d$ since we prefer the transform to be expressed as a rotation, a scaling, and an XY offset.  
Also, we add size and location constraints to fix the first image. $a = e = 1; b= d = c = f = 0.$
\end{itemize}

The we create and solve the normal equations. 
With no correction, we observe the shrinking as expected (see Table 1).  
We then correct for this by using the following intuition.  
Although the scale quite considerably from one end of the stack to the other, the angles between each adjacent plane pair can seek their own best value, with no restoring force to move them from the optimum.
Assuming the angle is approximately right, even if the scale is not, we can then add a constraint that encourages the scale to 1, at the angle each tile had in the initial solution.  
(If the derived angle is $\theta$, then we require $a = e = \cos(\theta)$ and $-b = d = \sin(\theta)$ in the least square sense.)  
In practice this gives a solution without scale shrinkage, and experiments as shown in Table 1 indicate this does not adversely affect the quality of the fit.

\providecommand{\e}[1]{\ensuremath{\times 10^{#1}}}
\begin{table}
\caption{Shrinkage experiments on synthetic data sets.  Ideally, the residual after correction, shown in the last column, should remain constant independent of the number of layers.}
\rule{0pt}{10ex} 
\begin{tabular}{ccccc}
   &  &  &   &  \\
number of  & predicted shrink & predicted shrink & actual     & residual\\
   layers  &  (analytic)      & (series)         &  shrinkage & after   \\
           &                  &                  &            & rescaling \\ \hline
          100    & 0.966    & 0.9681 &           0.9533 & 4.04 px  \\
          200    & 0.865    & 0.8803 &           0.8512 & 4.03 px \\
          500    & 0.156    & 0.5095 &           0.4464 & 4.05 px \\
         1000    & -        & 0.1486 &           0.1076 & 4.04 px \\
         2000    & -        & $1.11\e{-2}$  & $6.09\e{-3}$ & 4.04 px \\
         5000    & -        & $4.58\e{-6}$  & $1.04\e{-6}$ & 4.02 px \\
        10000    & -        & $1.05\e{-11}$ & $1.17\e{-11}$ & 4.44 px
\end{tabular}
\end{table}

\end{document}